\documentclass[aps,reprint,prr,longbibliography,a4paper,twocolumn,superscriptaddress,citeautoscript]{revtex4-1}
\usepackage[normalem]{ulem}
\usepackage[svgnames]{xcolor}
\usepackage[colorlinks=True,linkcolor=DarkRed,citecolor=ForestGreen,urlcolor=MediumBlue]{hyperref}
\usepackage[english]{babel}
\usepackage{lmodern}
\usepackage[T1]{fontenc}
\usepackage{amsmath}
\usepackage{units}
\usepackage[pdftex]{graphicx}
\usepackage{grffile}
\usepackage{color}
\usepackage[bottom]{footmisc}
\usepackage{graphicx}
\usepackage{dcolumn}
\usepackage{bm}\let\vec\bm 
\usepackage{soul}
\usepackage{amsfonts}
 \usepackage{braket}
\begin{document}%
\title{Superconductivity in metallic hydrogen}
\author{Dirk~van~der~Marel$^{1*}$ and Christophe~Berthod$^{2*}$
\vspace{1\baselineskip}
\\
$^{1,2}$Department of Quantum Mater Physics, University of Geneva, 24 Quai Ernest-Ansermet, 1211 Geneva 4, Switzerland
\\
$^{1}$Lead contact
\\
$^*$Correspondence: dirk.vandermarel@unige.ch
\\
$^{**}$Correspondence: christophe.berthod@unige.ch
\vspace{1\baselineskip}
\\
This manuscript version is made available under the CC-BY-4.0 license 
https://creativecommons.org/licenses/by/4.0/
\vspace{1\baselineskip}
}
\date{20 November 2024}
\maketitle

\section*{SUMMARY}

Superconductivity, the lossless flow of electric current, occurs typically at very low temperatures. 
A possible exception is highly pressurized hydrogen, for which room temperature superconductivity has been predicted.
However, as a result of various approximations used, conflicting theoretical predictions exist for the temperatures where superconductivity is expected to occur in highly pressurized hydrogen.
Here we avoid those approximations and exploit the ``jellium'' model proposed in 1966 by De Gennes, where superconductivity involves the combination of Coulomb repulsion between the electrons and Coulomb attraction between the protons and the electrons.
We confirm that metallic hydrogen should indeed exhibit superconductivity, but this is limited to temperatures far below previous estimates. 
We also find that the superconducting order develops over an energy range significantly exceeding the characteristic phonon energy, and that the phase of the order parameter flips 180 degrees at the characteristic phonon energy above and below the Fermi energy.

\noindent

\section*{KEYWORDS}


superconductivity, BCS theory, jellium model, hydrogen, Coulomb repulsion, dielectric function, electron-phonon coupling, isotope effect, Jupiter, white dwarf

\section*{INTRODUCTION}

According to the theory of Bardeen Cooper and Schrieffer~\cite{bardeen1957}, superconductivity is caused by electron-phonon coupling, the coupling of the electrons to vibrations of the lattice. A pedagogical description of this mechanism was introduced by De Gennes using the ``jellium model'', demonstrating that electron-phonon coupling can result in an effective interaction between electrons which overcomes the Coulomb repulsion between the electrons~\cite{degennes1966}. De Gennes indicated a number of restrictions for the range of applicability of the jellium model, notably that the model assumes the absence of core electrons. This limits the part of the periodic table where the model can be applied to hydrogen and helium, of which hydrogen is by far the most promising candidate. 
In 1968, Ashcroft suggested that the bulk of Jupiter is composed of hydrogen in the metallic state and that part of the bulk (with a temperature 100--200~K) may be in a superconducting state~\cite{ashcroft1968}. Various refinements were made in later papers using Eliashberg strong coupling theory~\cite{richardson1997,barbee1989,ashcroft2004}, predicting superconductivity above 200~K. In 1981, Jaffe and Ashcroft pointed out the possibility that metallic hydrogen may be a superconducting liquid in the density range $3.9\times10^{23}~\mathrm{cm}^{-3} <n<7.3\times10^{23}~\mathrm{cm}^{-3}$, with a maximum critical temperature $T_c\approx141$~K for $n=6.4\times10^{23}~\mathrm{cm}^{-3}$. 
The possibility that pulsars are cold magnetic white dwarfs was discussed, also in 1968, by Ginzburg and Khirzhnits~\cite{ginzburg1968}. They pointed out that the superconductivity of a certain layer of the star may be an essential factor and derived a simple expression for $T_c$ for light elements. Their expression for hydrogen, based on De Gennes' jellium model, is reproduced in Eq.~(\ref{eq:Ginzburg}). In a subsequent paper, Khirzhnits~\cite{kirzhnits1969} published a slightly different version of this expression, which is reproduced in Eq.~(\ref{eq:kirzhnits}).
In recent years, superconductivity in hydrogen-rich compounds has seen a revival, where numerical predictions and experimental confirmation of superconductivity near and above room temperature have entered the stage~\cite{boeri2022}---and to some extent left it again~\cite{garisto2024}. The calculation of $T_c$ for a multi-element material is a complex task, and a certain degree of coarse-graining of the momentum dependence and energy dependence of the electron-phonon coupling and the screened Coulomb interaction is unavoidable.

\begin{figure}
\includegraphics[width=\columnwidth]{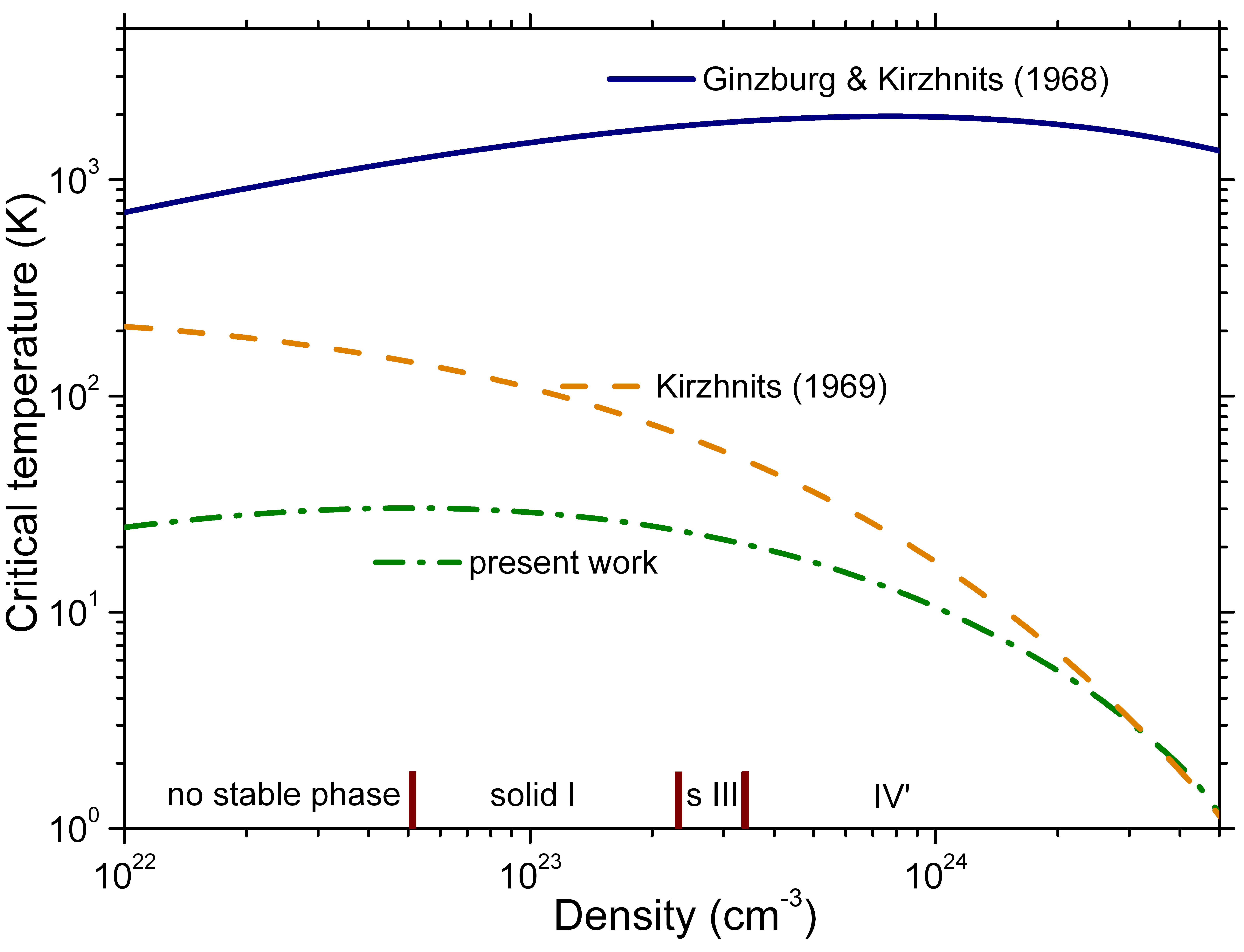}
\caption{Superconducting transition temperature of metallic hydrogen. 
Comparison of three calculations of $T_c$ in hydrogen at different densities, each of them using the jellium model~\cite{degennes1966}. Solid blue curve: Ginzburg and Kirzhnits expression~\cite{ginzburg1968} reproduced in Eq.~(\ref{eq:Ginzburg}). 
Dash-dotted orange curve: Kirzhnits' expression~\cite{kirzhnits1969} reproduced in Eq.~(\ref{eq:kirzhnits}).
Dashed green curve: present result.
The different phases  indicated along the bottom axis refer to the zero-temperature limit. The labels “solid I” (solid, phase I) and “s III” (solid, phase III) and IV' are adapted from Fig.~3 of Ref.~\cite{howie2015}, where IV' is possibly a liquid phase~\cite{eremets2011}. To relate the pressure to the density, the data in Fig.~4 of Ref.~\cite{mao1994} were used. In the region “no stable phase” the density is below that of solid molecular hydrogen in the limit of zero pressure~\cite{dewar1899}.
}
\end{figure}

The values of $T_c$ that follow from Ginzburg and Kirzhnits' expressions are displayed in Fig.~1. The fact that the results are so different may have to do with approximations beyond those of the jellium model,  that were necessary in view of limitations of the computational resources available at that time. For this reason, and without pretence of addressing the pairing mechanism of the aforementioned more complex materials, we return to the model interaction proposed by De Gennes and explore its properties within the BCS formalism~\cite{bardeen1957} without additional approximations. 

\section*{RESULTS}

Physically, a ``jellium'' would be realized if the material could be made dense enough, that not only the electrons but also the nuclei are in a liquid state. The formal requirement is that the Wigner-Seitz radius (radius of a sphere whose volume is equal to the volume per particle) of the electrons and that of the nuclei are both sufficiently small. For the case of hydrogen, this possibility was proposed---and its properties were explored---in a series of papers by Ashcroft and collaborators~\cite{jaffe1981,jaffe1983,moulopoulos1991,babaev2004} and by Tenney \textit{et al.}~\cite{tenney2021}. A recent review of the phases of highly compressed hydrogen up to 500 GPa (Carlo Pierleoni in Ref.~\cite{boeri2022}) does not address this proposal, but points out that the lack of experimental information about the crystalline structure together with the large nuclear quantum effects for hydrogen poses a challenge to \textit{ab initio} calculations. The jellium model is an interesting alternative approach since (i) crystallographic information is not required, (ii) the mathematics is relatively simple, (iii) it is optimally suited for hydrogen because the model assumes the absence of core electrons, and (iv) the model rests on only two parameters: the Wigner-Seitz radius of the electrons and the mass of the nuclei. 

\begin{figure}
\includegraphics[width=\columnwidth]{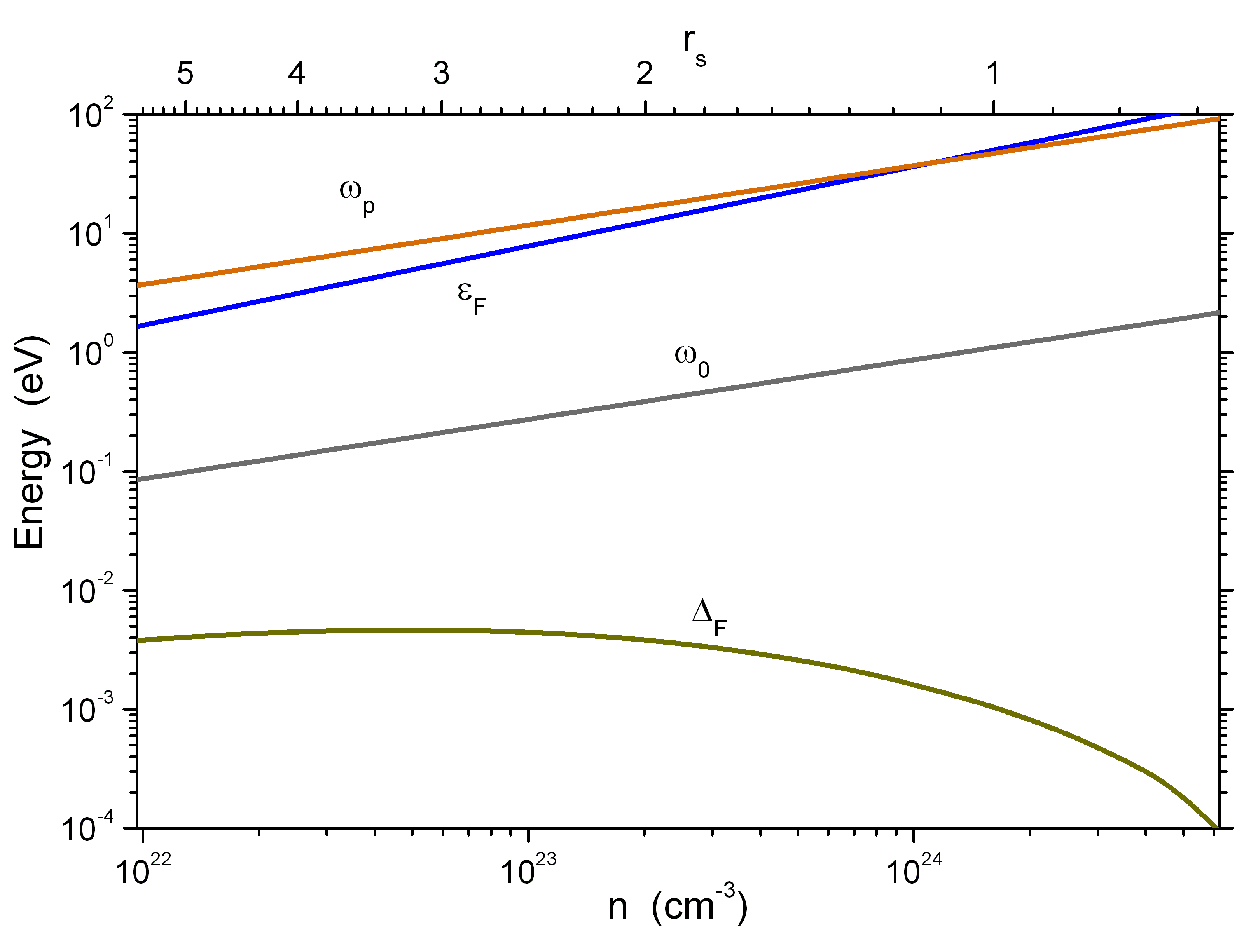}
\caption{Relevant energy scales of metallic hydrogen as a function of density.
The lower (upper) horizontal axis indicates the density (Wigner-Seitz radius).
}
\end{figure}

At this point, it is useful to introduce the relevant energy scales of the system. In Fig.~2, the plasmon energy of the electrons $\omega_p$, the Fermi energy of the electrons $\varepsilon_{\mathrm{F}}$, the plasmon energy of the nuclei $\omega_0$, and the gap energy at the Fermi surface $\Delta_{\mathrm{F}}$ (anticipating for the latter the discussion of the present manuscript) are displayed as a function of the electron density $n$ and---on the opposite axis--- the Wigner-Seitz radius $r_s=a_0^{-1}(4\pi n /3)^{-1/3}$, where $a_0=0.529177$~\AA{} is the Bohr radius. For the density range relevant for metallic hydrogen ($r_s\sim 1.5$) the plasma energy and the Fermi energy are of the same order of magnitude, and for all densities $\omega_p, \varepsilon_{\mathrm{F}}\gg\omega_0\gg\Delta_{\mathrm{F}}$. In Methods - Conventional Approach it is shown that the electron-phonon coupling parameter $\lambda$ is smaller than 0.5 for all values of $r_s$, which is in the weak coupling regime. Furthermore, since $\omega_0\ll \varepsilon_{\mathrm{F}}$, the system is in the regime where the motion of the nuclear particles is slow compared to that of the electrons. All in all, this implies that for the density range of interest the BCS model~\cite{bardeen1957} can in principle be used. 

Following De Gennes~\cite{degennes1966}, we adopt as a starting point the Coulomb interaction screened by electrons and nuclear particles $V^{\mathrm{eff}}$ as:
\begin{equation}
V^{\mathrm{eff}}(q,\omega) =\frac{4 \pi e^2}{q^2}
\frac{1}{\epsilon(q,\omega)},
\label{eq:1}
\end{equation}
where $e$ is the electron charge, $\omega$ is the angular frequency, $\epsilon(q,\omega)$ is the dielectric function and $q=|\vec{q}|$. Here $\hbar\vec{q}=\hbar(\vec{k}-\vec{p})$ represents the transferred momentum in a process where  two electrons in the plane wave state $\ket{\vec{k},-\vec{k}}$ scatter to $\ket{\vec{p},-\vec{p}}$
and $\hbar$ is the Planck constant. We furthermore adopt the procedure of De Gennes~\cite{degennes1966} and Kirzhnits~\cite{kirzhnits1969}, and substitute $\hbar\omega=\varepsilon_k-\varepsilon_p$. The rational, provided on pp.~100--102 of Ref.~\cite{degennes1966}, is that with this substitution the effective interaction has exactly the form of the matrix element $V_{k,p}$ between an initial state $\ket{\vec{k},-\vec{k}}$ and a final state $\ket{\vec{p},-\vec{p}}$.  

For a high-density electron gas, the dielectric function in the jellium approximation is described by 
\begin{equation}
\epsilon(q,\omega)= 1+\frac{k_{0}^2}{q^2}-\frac{\omega_0^2}{\omega^2},
\label{eq:2}
\end{equation}
where $k_{0}$ is the Thomas-Fermi screening wavevector and $\omega_0$ is the plasma frequency of the positively charged nuclei. 
Expressing the electron energies and the gap energy in units of the Fermi energy, \textit{i.e.}, $\varepsilon_k=x\varepsilon_{\mathrm{F}}$, $\Delta_k=\psi_x\varepsilon_{\mathrm{F}}$, 
the gap equation obtains the following compact form (Methods - Gap equation)

\begin{align}
&\psi_x=-\int_{0}^{\infty}dy\,\sqrt{y}\upsilon^s_{x,y}
\frac{\psi_y}{2\sqrt{(y-1)^2+\psi_y^2}} \times
\nonumber\\
&\tanh\left(\frac{\sqrt{(y-1)^2+\psi_y^2}}{2\tau}\right),
\end{align}
where 
\begin{align}
&\upsilon^s_{x,y}
= \frac{(x-y)^2}{2 \sqrt{xy}[4{z}_{0}^2-(x-y)^2\gamma^{-2}]}  \times
\nonumber\\
&\ln\left|\frac{ {4(x-y)^2 -   \left(\sqrt{x} -\sqrt{y}\right)^2 [4{z}_{0}^2-(x-y)^2\gamma^{-2}]} }
{ {4(x-y)^2 - \left(\sqrt{x} +\sqrt{y}\right)^2 [4{z}_{0}^2- (x-y)^2\gamma^{-2}]} }\right|
\label{eq:upsilon_kernel}
\end{align}
is the interaction in the $s$-wave pairing channel, and is displayed in the top panel of Fig.~3.
\begin{figure}
\includegraphics[width=\columnwidth]{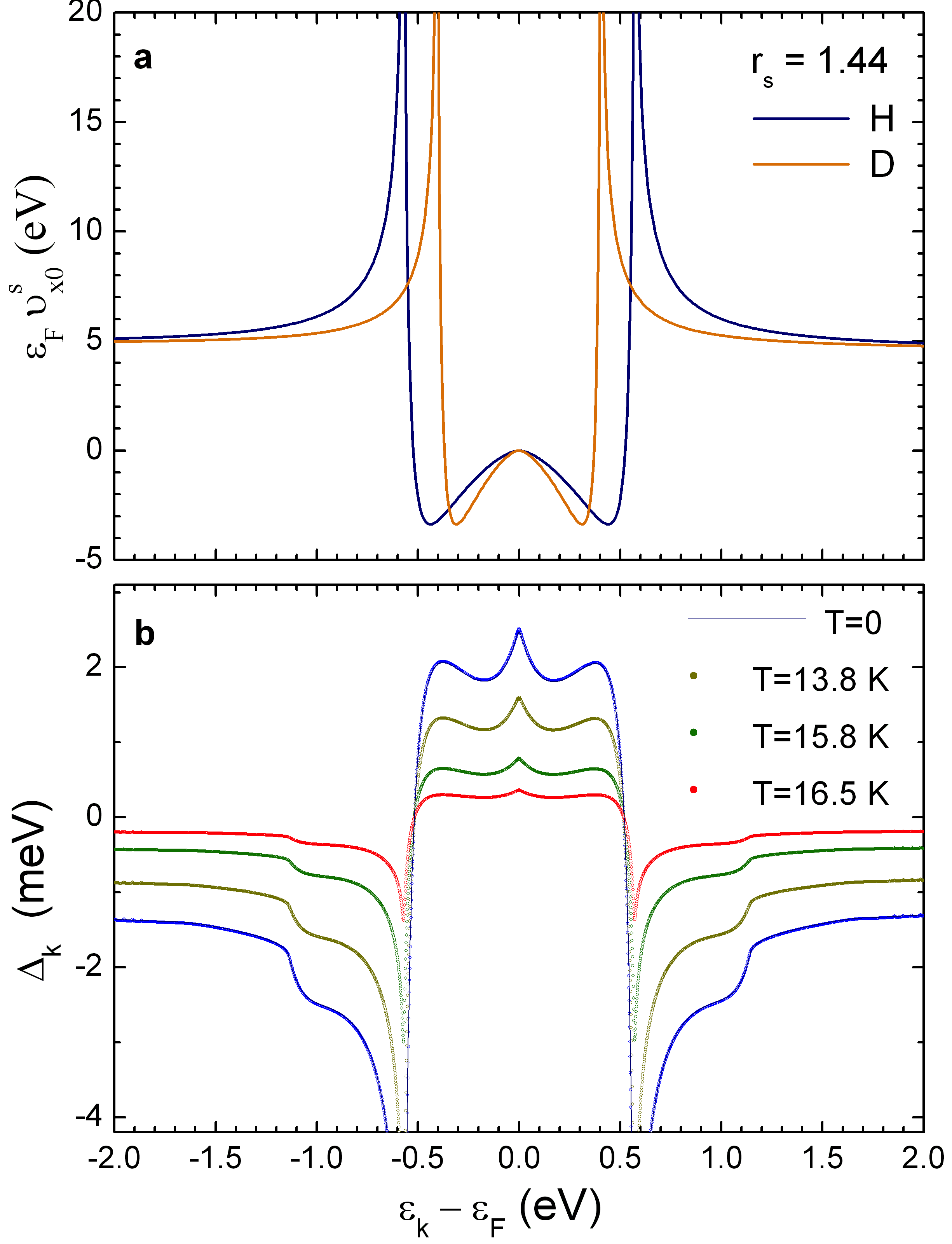}
\caption{Energy dependence of the interaction and the gap function. {\bf a} s-wave coupling function, $\varepsilon_{\mathrm{F}}\upsilon^s_{x,0}$, as a function of energy $\varepsilon_k=\varepsilon_{\mathrm{F}}x$ in the case of hydrogen ($m_N=m_p$, blue) and deuterium ($m_N=2m_p$, orange) with $r_s=1.44$. {\bf b}  Energy dependence of the order parameter for $r_s=1.44$ in the case of hydrogen ($m_N=m_p$), and for four temperatures below $T_c$~($\sim 17$~K).
For $T=0$ the results computed with two different numerical codes are compared. Open symbols : Numerical code where the integrals are calculated with the trapezoidal rule (Methods - Numerical implementation). Continuous curve: Numerical code where the integrals are calculated with a piecewise-continuous function of $x$ (Methods - Solution without discretization). The difference of these two curves is given in Fig. 7.
}
\end{figure}
The two constants in this expression,
\begin{equation}
z_0^2=\frac{4m_e}{3m_N}, \qquad
\gamma^2=\left(\frac{2}{3\pi^2}\right)^{2/3}r_s,
\end{equation}
depend uniquely on the nuclear mass $m_N$ of the isotope (hydrogen or deuterium in the present context) and the Wigner-Seitz radius $r_s$.
Remarkably, for zero energy shift ($x=y$) the interaction is exactly zero. The interaction is furthermore negative for small energy shift, passes through a minimum, changes sign, passes through a positive logarithmic singularity and asymptotically approaches the Thomas-Fermi screened Coulomb repulsion at high energies. In the most common implementation of the theory of Bardeen Cooper and Schrieffer~\cite{bardeen1957} the interaction is assumed to be constant and attractive in the interval $\pm\omega_0$ around the Fermi energy, corresponding to an interaction at the Fermi level that is both non-zero and attractive. Given that in the present model the interaction at the Fermi energy is zero, it is not immediately obvious that such an interaction would induce the Cooper instability~\cite{cooper1956}. One of the motivations for the present study was to investigate whether the BCS variational wave function has a non-trivial solution with this interaction. To address this question, we solved the gap equation numerically (see Methods - Numerical implementation) and calulated the gap function, the condensation energy (Eq.~\ref{eq:econd}), and the specific heat (Eq.~\ref{eq:CV}).
Instead of substituting $\hbar\omega=\varepsilon_k-\varepsilon_p$ in the interaction and solving the BCS gap equations, it is in principle possible to solve the Eliashberg equations and calculate $T_c$~\cite{eliashberg1960}. While the relevant set of equations for the present case could be identified without a problem, due to the simultaneous energy- \emph{and} momentum dependence of the interaction, the numerical algorithm poses a computational challenge which we haven't been able to overcome yet.

The resulting gap function is shown in the lower panel of Fig.~3 for hydrogen assuming a Wigner-Seitz radius $r_s=1.44$ at four temperatures below $T_c$. We see that the gap function changes sign near the energy where $\upsilon^s_{x,0}$ changes sign, and shows a negative peak (which is however not a singularity) near the energy where $\upsilon^s_{x,0}$ has a log-singularity. The first discussion of this 180 degrees phase shift was by Bogoliubov, Tolmachov and Shirkov~\cite{bogoljubov1958} and it was illustrated in Fig.~2 of Ref.~\cite{morel1962} and Fig.~1 of Ref.~\cite{pereg1992}.
One of the remarkable features is that the region of finite order parameter---and temperature evolution thereof---is not limited in any way to a narrow range around the Fermi energy. Although this doesn't contribute significantly to the free energy of the electrons, experimental probes such as photo-emission, optics and tunneling spectroscopy can in principle pick up the temperature dependency of the order parameter at energies far beyond the characteristic energy of the phonons that mediate the pairing interaction.
Another feature---not visible on this scale---is that for large positive energy $\Delta_k=\varepsilon_{\mathrm{F}}\psi_x$ is proportional to $1/x$. 

\begin{figure}
\noindent\includegraphics[width=\columnwidth]{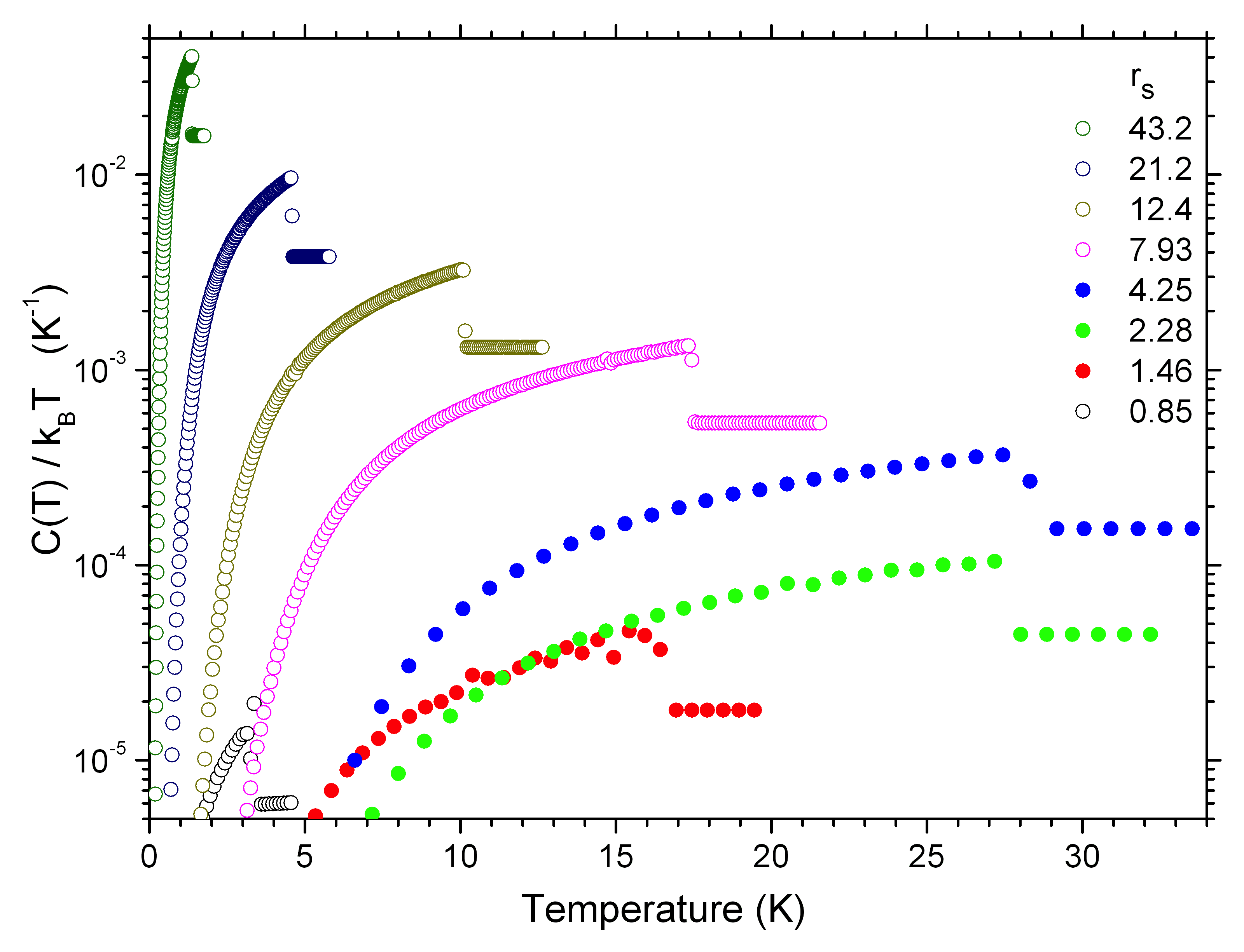}
\caption{Specific heat  divided by temperature for eight different values of $r_s$ in the case of hydrogen ($m_N=m_p$).
The number of iterations of the gap equation to full convergence depends on $r_s$ and on temperature and increases when the order parameter shrinks. For practical reasons, the iterative process was terminated after 5000 iterations, even if close to $T_c$ full convergence wasn’t reached yet. Incomplete convergence is revealed as a broadening of the specific heat jump.
The results shown for $r_s\leq 4.25$ were binned in intervals averaging over 5 adjacent temperatures to reduce jitter. 
}
\end{figure}
The temperature-dependent specific heat  is shown in Fig.~4. The drop in the specific heat was used to determine $T_c$, shown in Fig.~5. The maximum is reached for $r_s\sim3$, where in the case of hydrogen $T_c\sim 30$~K. To obtain an accurate estimate of $T_c$, the gap equation has to be solved for many temperatures for each $r_s$ value. Since this is a time-consuming calculation, we also tested a less cumbersome method, namely to use the following proxy for $T_c$ ($k_{\mathrm{B}}$ is the Boltzmann constant):
\begin{equation}
T_c^* = A \sqrt{\varepsilon_{\mathrm{F}}E_{c}} /k_{\mathrm{B}}.
\label{eq:tstar}
\end{equation}
In the most commonly used implementation of BCS theory, the interaction is assumed constant between $\pm \omega_0$ and zero elsewhere. In this case the condensation energy \emph{per particle} is
\begin{equation}
E_{c}= \frac{3 \Delta_0^2}{8\varepsilon_{\mathrm{F}}} 
\end{equation}
and the gap over $T_c$ ratio
\begin{equation}
\frac{2\Delta_{0}}{k_{\mathrm{B}}T_c}= 3.53.
\end{equation}
\begin{figure}
\noindent\includegraphics[width=\columnwidth]{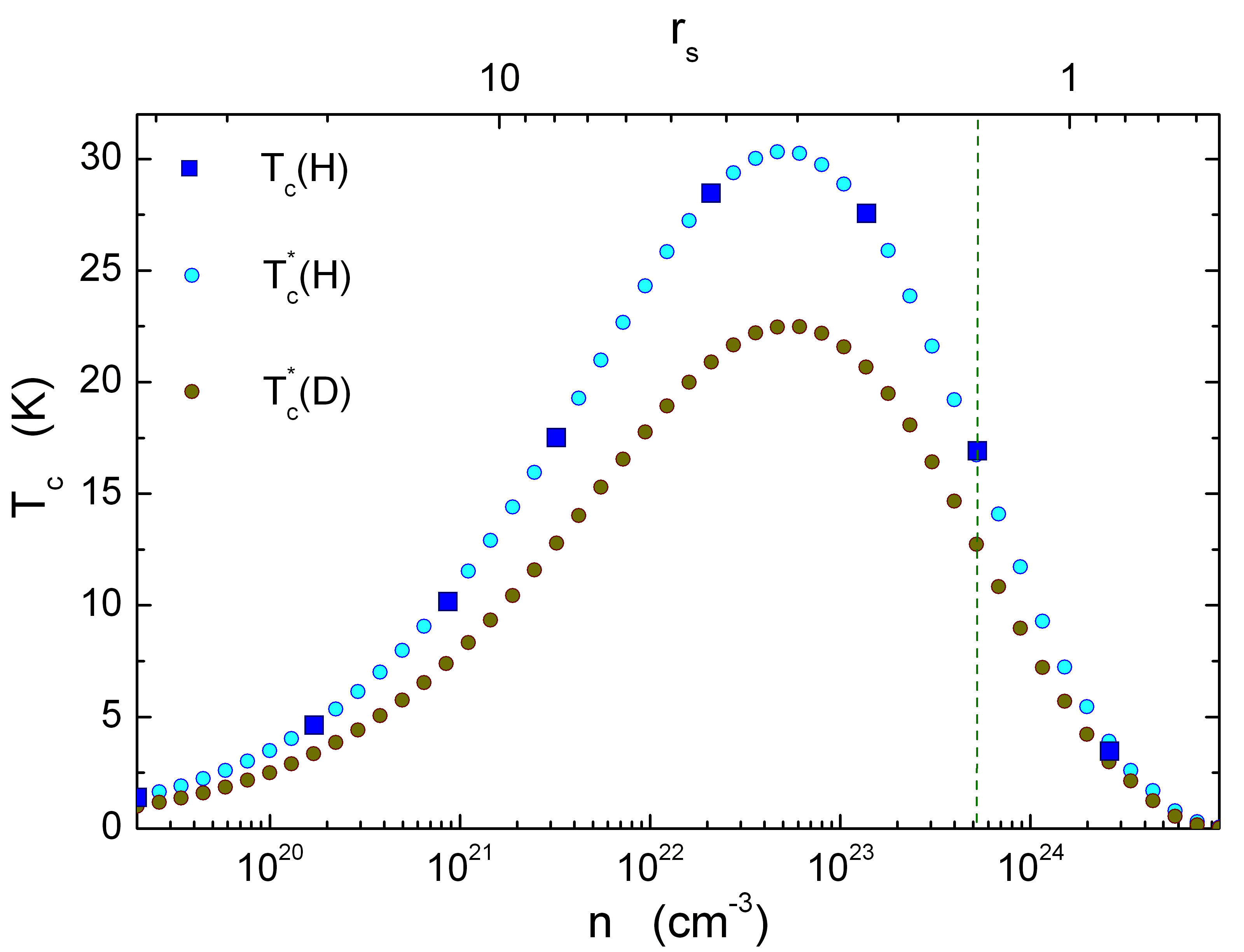}
\caption{Superconducting transition temperature as a function of density.
Square symbols: $T_c$ for $m_N=m_p$ (hydrogen) determined from the specific heat jump (Fig.~4). The vertical dashed line indicates the density of metallic hydrogen~\cite{mazin1995}. 
Open symbols: Effective $T_c$ determined from the condensation energy [Eq.~(\ref{eq:tstar}) with $A=0.925$] for hydrogen (blue) and deuterium (orange, $m_N=2m_p$).
}
\end{figure}
Combining these two expressions one obtains $A=0.925$. 
The results for $T_c$ and $T_c^*$ are displayed in Fig.~5 for hydrogen (blue symbols) and deuterium (orange symbols). 
The density dependence of $T_c^*$, shown in Fig.~5, is described by a dome. The result for hydrogen is also displayed as the dashed green curve in Fig.~1.
For the  $T_c$'s of Fig.~5, using $\alpha=\ln (T_c^{\mathrm{H}}/T_c^{\mathrm{D}})/\ln(2)$ we found that $\alpha$ gradually decreases from 0.48 for $n=10^{20}$~cm$^{-3}$ to 0.35 for $n=10^{24}$~cm$^{-3}$ (see Fig.~8).
Without Coulomb repulsion the isotope effect on $T_c$ is described by $T_c\propto m_N^{-\alpha}$ where $\alpha=0.5$. This value changes if the Coulomb repulsion is taken into account~\cite{morel1962}, as demonstrated in Fig.~8. 

\section*{DISCUSSION}
Ashcroft estimated $T_c\sim 200$~K for dense hydrogen~\cite{ashcroft1968,richardson1997} assuming a coupling constant $\lambda\approx 1.5$. The $T_c$ values obtained with the jellium model are an order of magnitude below these estimates as a result of the relatively low coupling constant which, as demonstrated in Methods - Conventional Approach, stays below $0.5$ for all densities.
A natural question is what the value of $T_c$ would become if, instead of using the jellium approximation, the full potential of the positively charged nuclear particles would be taken into account. Since no structural information is available for densities corresponding to the metallic phase of hydrogen, it is not possible at the present stage to predict with full confidence the superconducting properties for the actual atomic configuration of solid or liquid metallic hydrogen beyond the jellium approximation. 
It is possible that the lattice potential plays a crucial role in causing strong electron-phonon coupling, an element that is inherently absent from the jellium model. 
Using \textit{ab initio} predictions for the crystal structure, McMahon and Ceperley~\cite{mcmahon2011} and Dogan \textit{et al.}~\cite{dogan2024} calculated $T_c$ values ranging from nearly zero to ~450 K for different phases of solid hydrogen up to 3.5 TPa. Tenney \textit{et al.}~\cite{tenney2021} investigated liquid superconductivity in metallic hydrogen by first-principles simulations and solving the Eliashberg equations, finding a pressure dependent dome with a maximum $T_c$ of 360~K for 780~GPa.

Hirsch~\cite{hirsch-book} has argued on the basis of Alfven’s theorem~\cite{alfven1942} that for the Meissner effect to occur it is essential that the charge carriers have \emph{negative} effective mass; otherwise the transition would not be reversible. On the one hand we have demonstrated in the present study that, with De Gennes' effective electron-electron interaction, the BCS variational wave-function of free electrons---which have a positive mass---has a non-trivial ground state with a finite order parameter. On the other hand, according to Hirschs' argument there can be no Meissner effect in this model in view of the positive mass. If this argument is correct, this would imply that interacting electrons \emph{can} condense into a state with a BCS-type order parameter but, in view of the positive mass of the electrons this state of matter differs in a fundamental way from a superconductor and does not have a Meissner effect. We nonetheless continue to use the terminology of superconductivity for the remainder of the discussion and return to the questions about superconductivity in white dwarf stars and in the interior of Jupiter mentioned in the introduction. To begin with the latter, the black-body temperature provides a lower limit of the temperature of a planet, which in the case of Jupiter is 109.9~K~\cite{williams2021}. For white dwarfs the situation is more complicated. For the coldest one that has been observed an upper limit of 3000~K was estimated~\cite{kaplan2014}. This does not exclude that the temperature is lower than 30~K, however, given the current age of the known universe it is unlikely that sufficient time has elapsed for a white dwarf to cool down to such a low temperature. If one accepts the jellium model that was originally proposed to argue that white dwarfs could be superconducting \cite{ginzburg1968}, it seems unlikely both for the case of Jupiter or for white dwarfs that the temperature is low enough (lower than 30~K) to contain metallic hydrogen in the superconducting state.

In conclusion, the superconducting pairing in metallic hydrogen involves an intricate interplay of the Coulomb repulsion between the electrons and an attractive phonon-mediated pairing mechanism. One of the consequences is that the superconducting order parameter has a 180$^\circ$ phase shift at the characteristic energy of the phonons~\cite{bogoljubov1958}. The outcome for the superconducting critical temperature ($T_c<30$~K) is much lower than the $T_c$ predicted in Refs.~\cite{ashcroft1968,richardson1997,barbee1989,ashcroft2004} and previous results using the same model~\cite{ginzburg1968,kirzhnits1969}. For energies far above the phonon energy range, the amplitude of the order parameter is of the same order of magnitude as that at the Fermi surface.

\subsection*{Limitations of the study}
Given the strong tendency of hydrogen to form dimers the electron-lattice coupling is likely an important factor. This factor is not taken into account in the “jellium” model used in this study. If the electron-phonon coupling is strong, it becomes necessary to use the Migdal-Eliashberg formalism instead of the BCS gap equations.  

\onecolumngrid
\section*{METHODS}

\subsection*{Resource availability}


\subsubsection*{Lead contact}


Requests for further information and resources should be directed to and will be fulfilled by the lead contact, Dirk van der Marel (dirk.vandermarel@unige.ch).

\subsubsection*{Materials availability}


This study did not generate new materials.

\subsubsection*{Data and code availability}


\begin{itemize}
    \item The theoretical data generated in this study are available in Ref.~\cite{opendataHOD}. These will be preserved for 10 years. All other data that support the plots within this paper and other findings of this study are available from the corresponding author upon reasonable request.
    \item The custom computer codes used to generate the results reported in this paper are available in Ref.~\cite{opendataHOD}.
    \item Any additional information required to reanalyze the data reported in this paper is available from the lead contact upon request.    
    
\end{itemize}


\subsection*{Previous expressions for $T_c$ using the jellium model}

Ginzburg and Kirzhnits~\cite{ginzburg1968} used the following expressions
\begin{align}
T_c=\hbar\omega_0 \exp{\left(-\frac{1}{NV}\right)}
\quad\text{with}\quad
NV =\frac{e^2}{\hbar v_{\mathrm{F}}} = \left(\frac{4}{9\pi}\right)^{1/3}r_s
\quad\text{and}\quad
\omega_0^2=\frac{4\pi n e^2}{m_p}.
\end{align}
The combination of these expressions gives
\begin{equation}
k_{\mathrm{B}} T_c =  
\mbox{Hr}\sqrt{\frac{3m_e}{m_N}}\frac{1}{r_s^{3/2}}
\exp{\left[-\left(\frac{9\pi}{4}\right)^{1/3}\frac{1}{r_s}\right]}.
\label{eq:Ginzburg}
\end{equation}
where Hr $\approx 27$~eV is the Hartree energy.
Kirzhnits~\cite{kirzhnits1969} used 
\begin{align}
\Delta=\alpha \omega_0 \exp{\left(-\frac{8}{\pi^2\alpha}\right)}
\quad\text{with}\quad
\alpha=\frac{e^2}{\pi v_{\mathrm{F}}} = \left(\frac{2}{3\pi^2}\right)^{2/3}r_s
\label{eq:kirzhnits_coupling}
\end{align}
and the same definition of $\omega_0$, stating that the critical temperature $T_c$ is connected with this quantity in the usual manner.
Combining these expressions gives
\begin{equation}
k_{\mathrm{B}}T_c=
\mbox{Hr} \frac{e^\gamma}{\pi}
\left(\frac{2}{3\pi^2}\right)^{2/3}
 \sqrt{\frac{3 m_e}{m_N} } \frac{1}{r_s^{1/2}}
\exp{\left[-2\left(\frac{12}{\pi}\right)^{2/3}\frac{1}{r_s}\right]}
\label{eq:kirzhnits}
\end{equation}
where $\gamma\approx 0.577$ is the Euler constant.

\subsection*{Gap equation}
\label{section:gap}

If the BCS variational wave function is applied to a system of band electrons with energy-momentum dispersion $\varepsilon_k$ and an electron-electron interaction $V^{\mathrm{eff}}_{\vec{k},\vec{p}}$, one obtains the following expression for the grand potential~\cite{rickayzen1965}
\begin{align}
\Omega&=
\sum_{\vec{k}}
\left[
-2k_{\mathrm{B}}T\ln\left(1+e^{-E_k/k_{\mathrm{B}}T}\right)
+\xi_k -E_k
+\frac{|\Delta_k|^2}{E_k}\tanh\left(\frac{E_k}{2 k_{\mathrm{B}}T}\right)
\right]+
\sum_{\vec{k},\vec{p}}
\frac{\Delta_k \Delta_p^*V^{\mathrm{eff}}_{\vec{k},\vec{p}}}{4E_kE_p\nu} 
\tanh\left(\frac{E_p}{2 k_{\mathrm{B}}T}\right)
\label{eq:free1}
\end{align}
where $\xi_k=\varepsilon_k-\mu$, $\mu$ is the chemical potential, and $E_k=\sqrt{\xi_k^2+|\Delta_k|^2}$. 
In equilibrium, where $\Delta_k$ corresponds to the set of values that minimizes $\Omega$,  the number of electrons follows from the relation 
\begin{equation}
N=-\frac{\partial \Omega}{\partial \mu}
=\sum_{\vec{k}}
\left[
1-\frac{\xi_k}{E_k}\tanh\left(\frac{E_k}{2 k_{\mathrm{B}}T}\right)
\right]
\label{eq:Ne}
\end{equation}
We reserve the symbols $N_0$ for the number of electrons, $n_0$ for the electron density and $\varepsilon_{\mathrm{F}}$ for the chemical potential at $T=0$ and in the absence of superconducting order.
$V^{\mathrm{eff}}_{k,p}$ represents the effective interaction between the electrons. Here we consider the screened Coulomb interaction~\cite{degennes1966}
\begin{equation}
V^{\mathrm{eff}}(q,\omega) =\frac{4 \pi e^2}{q^2}
\frac{1}{\epsilon(q,\omega)} 
\end{equation}
For $\epsilon(q,\omega)$ we adopt the the jellium model:
\begin{equation}
\epsilon(q,\omega)= 1+\frac{k_{0}^2}{q^2}-\frac{\omega_0^2}{\omega^2} 
\label{eq:epsilon_q_omega}
\end{equation}
where
\begin{equation}
k_{0}^2={\frac{4e^2m_ek_{\mathrm{F}}}{\pi \hbar^2}}, 
\hspace{1cm}
\omega_0^2={\frac{4\pi n e^2}{m_N}}
\label{eq:k0_omega0}
\end{equation}
so that
\begin{equation}
V^{\mathrm{eff}}(q,\omega)  =
\frac{4 \pi e^2\omega^2}{k_{0}^2\omega^2 +(\omega^2-\omega_0^2)q^2} 
\label{eq:Uscr2}
\end{equation}
The energy dispersion of the longitudinal phonons follows from the condition $\epsilon(q,\omega)=0$. These expressions are valid in the limit of linear response and require that the vibrations are purely harmonic. To obtain a more compact formulation we multiply both sides of Eq.~(\ref{eq:free1}) with $1/(N_0\varepsilon_{\mathrm{F}}) = 3\pi^2 /(\nu k_{\mathrm{F}}^3 \varepsilon_{\mathrm{F}})$, divide the electron energies by the Fermi energy and adopt Ashcroft's notation~\cite{ashcroft2004} $\gamma$ for the Thomas-Fermi factor. Concretely we define the following set of dimensionless quantities
\begin{equation}\begin{aligned}
w&=\frac{\Omega}{N_0\varepsilon_{\mathrm{F}}},
&x&=\frac{\varepsilon_k}{\varepsilon_{\mathrm{F}}},
&y&=\frac{\varepsilon_p}{\varepsilon_{\mathrm{F}}},
&z&=x-y=\frac{\hbar\omega}{\varepsilon_{\mathrm{F}}}
\\
\psi_x&=\frac{\Delta_k}{\varepsilon_{\mathrm{F}}},
&\tau&=\frac{k_{\mathrm{B}}T}{\varepsilon_{\mathrm{F}}},
&z_0&=\sqrt{\frac{4m_e}{3m_N}},
&\gamma&=\frac{k_0}{2k_{\mathrm{F}}} =\left(\frac{2}{3\pi^2}\right)^{1/3}\sqrt{r_s}.
\end{aligned}\end{equation}
Since $\psi_k$ is isotropic for the s-wave pairing assumed here, $V^{\mathrm{eff}}_{\vec{k}\vec{p}}$ is the only term in the free-energy expression that depends on the angular coordinates of $\vec{k}$ and $\vec{p}$, and only on the relative angle of these two. This way we obtain
\begin{align}
w&=
\frac{3}{4}\int_{0}^{\infty} 
dx\,\sqrt{x}
\left[
-2\tau\ln\left(1+e^{-E_x/\tau}\right)
+\xi_x
-E_x 
+\frac{|\psi_x|^2}{E_x}\tanh\left(\frac{E_x}{2\tau}\right)
\right]+
\nonumber\\
&\quad+
\frac{3}{4}
\int_{0}^{\infty}dx\,
\sqrt{x}
\int_{0}^{\infty}dy\,
\sqrt{y}
 \frac{\psi_x \psi_y^* \upsilon^s_{x,y}}{4E_xE_y}
\tanh\left(\frac{E_x}{2\tau}\right) 
\tanh\left(\frac{E_y}{2\tau}\right),
\end{align}
where $\xi_x=x-\mu/\varepsilon_{\mathrm{F}}$, $E_x=\sqrt{\xi_x^2+|\psi_x|^2}$ and
\begin{equation}
\upsilon^s_{x,y}
= \frac{z^2}{2 \sqrt{xy}[4{z}_{0}^2-z^2\gamma^{-2}]}  
\ln\left|\frac{ {4z^2 -   \left(\sqrt{x} -\sqrt{y}\right)^2 [4{z}_{0}^2-z^2\gamma^{-2}]} }
{ {4z^2 - \left(\sqrt{x} +\sqrt{y}\right)^2 [4{z}_{0}^2- z^2\gamma^{-2}]} }\right|
\end{equation}
is the interaction in the $s$-wave pairing channel.
The equilibrium condition
\begin{equation}
\forall x:\qquad\frac{\partial w}{\partial{\psi_x}}=0
\end{equation}
results in the gap equation
\begin{equation}\label{eq:gap_equation}
\psi_x= - \int_{0}^{\infty}  
dy\,
\sqrt{y} 
\frac{\psi_y\upsilon^s_{x,y}}{2E_y} 
\tanh\left(\frac{E_y}{2\tau}\right).
\end{equation}
In a bulk 3-dimensional sample the macroscopic charge has the effect of maintaining the number of electrons inside the sample constant. Consequently the chemical potential is temperature dependent, which can be observed experimentally as a temperature dependence of the workfunction~\cite{rietveld1992}. To account for this effect, the condition~(\ref{eq:Ne}) must be applied simultaneously with the solution of the gap equation. Since the shift of chemical potential is of order $\Delta^2/4\varepsilon_{\mathrm{F}}$, the effect becomes relevant in superconductors where $\Delta$ is of comparable size as $\varepsilon_{\mathrm{F}}$ which, given the hierarchy of energy scales shown in Fig.~2, clearly doesn't apply to the system considered here. For this reason it makes no significant difference if the gap equation is solved in the canonical ensemble or in the grand canonical ensemble. Here we follow the latter approach and fix the chemical potential at $\mu=\varepsilon_{\mathrm{F}}$. 
\\
The entropy is
\begin{equation}
S(T)=-\frac{dw}{d\tau} k_{\mathrm{B}}
=\frac{3}{2}k_{\mathrm{B}}\int_0^{\infty} dx \sqrt{x}
\left[\frac{\ln({1+e^{E_x/\tau}} )}{1+e^{E_x/\tau}}  +\frac{ \ln ({1+e^{-E_x/\tau}})}{1+e^{-E_x/\tau}}\right] 
\end{equation}
The specific heat is obtained by (numerical) differentiation of $S(T)$
\begin{equation}
C(T)=T\frac{d}{dT}S(T)
\label{eq:CV}
\end{equation}
\begin{figure}
\includegraphics[width=0.5\columnwidth]{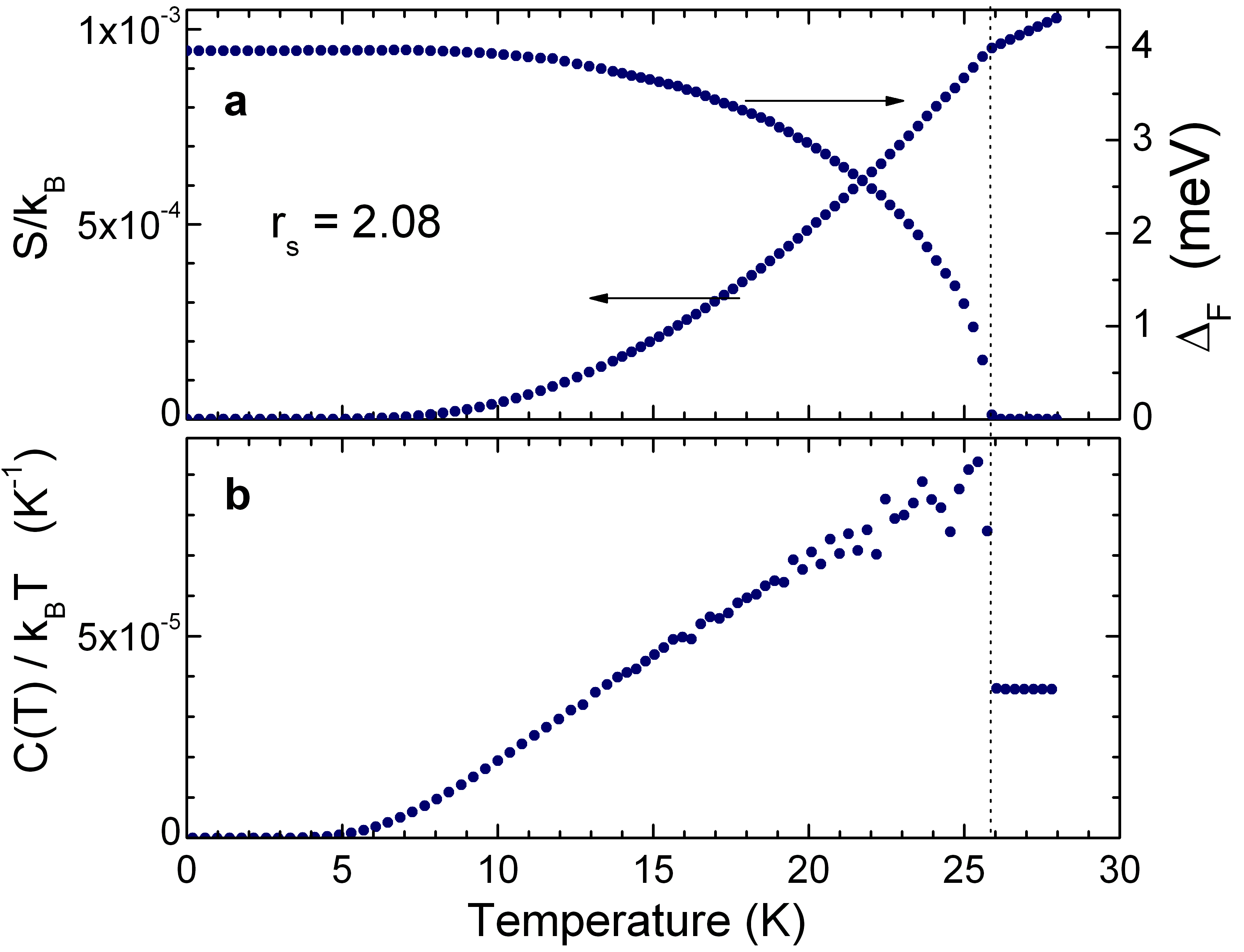}
\caption{Temperature dependence of the order paramater at the Fermi surface, the entropy and the specific heat.
These results are for $r_s=2.08$ in the case of hydrogen ($m_N=m_p$) calculated with the trapezoidal rule (“Numerical implementation” in Methods).  The noise of $C(T)/T$ (panel {\bf b}) is stronger than that of the entropy in panel {\bf a}, because it was calculated taking the numerical derivative of $S(T)$.
}
\end{figure}
In Fig.~6 entropy, specific heat and order parameter at the Fermi surface are displayed as a function of temperature for the case $r_s=2.08$.
The scatter of $C/T$ data is caused by the finite energy mesh and the automatic adjustment thereof at each successive temperature to accommodate the shrinking of the order parameter when approaching $T_c$.
\\
The present formalism is similar to the one treated by Swihart~\cite{swihart1962}, with a number of differences: we have a factor $1/\sqrt{x}$ in front of the interaction, Eq.~\ref{eq:upsilon_kernel}, which comes from the energy dependence of the electronic DOS; a factor depending independently on $x$ and $y$ corrects the second term inside the logarithm. There are further differences in the numerical solution method, as we do not introduce additional simplifications, while the solution proposed by Swihart uses a cutoff, assumes particle-hole symmetry of the gap function and replaces the interaction by a square well. Another treatment of similar nature is the one of Pereg, Weger and Kalma~\cite{pereg1992} who solved the energy dependence of the gap for a phonon-mediated interaction assuming momentum independent phonons without Coulomb interaction between the electrons.

\subsection*{Numerical implementation}
\label{section:Discrete}

\begin{table}[!!b!!]
\begin{tabular}{|c|c|}
\hline
\mbox{Energy interval}&\mbox{No. of equidistant energy values } \nonumber\\
\hline\hline
$-1.0<\xi_j<0.1$ & $n$  \nonumber\\
$ -0.1<\xi_j<-4\sqrt{\psi_F^2+\tau^2}$ & $4n$  \nonumber\\
$-4\sqrt{\psi_F^2+\tau^2} <\xi_j<4\sqrt{\psi_F^2+\tau^2}$ & $n$ \nonumber\\
$4\sqrt{\psi_F^2+\tau^2}<\xi_j<0.1$ & $4n$\nonumber\\
$0.1<\xi_j<1$ &  $n$\nonumber\\
$1<\xi_j<100$ & $n$\nonumber\\
$100<\xi_j<10^4$ &$n$ \nonumber\\
$10^4<\xi_j<\infty$ & \mbox{analytical continuation}\nonumber\\
\hline
\end{tabular}
\caption{\label{table1} {\bf Discretization of the energy axis} used in numerical integrals.}
\end{table}
It is easy to demonstrate that at $T_c$ the gap function is real, apart from an arbitrary phase that doesn't depend on $k$ and that we can set to zero. It is in principle possible that at an additional transition takes place to a state with a $k$-dependent phase. Since this requires a finite amplitude of the order parameter, it can only happen at a temperature lower than $T_c$ so that it wouldn't influence $T_c$. Since our purpose is to determine $T_c$, we can limit the present discussion to a real-valued order parameter. In this case the expression of the grand potential formulated for numerical coding is :
\begin{align}
w&=
\frac{3}{4}
\sum_h
\left[
-2\tau\ln\left(1+e^{-X_h/\tau}\right)
+ \xi_h  -X_h
+\frac{\psi_h(2\psi_h-\phi_h)}{2X_h}\tanh\left({X_h}/{2\tau}\right)
\right] D_h
\\
\mbox{where}&\nonumber
\\
\xi_h&= x_h-1
\\
X_h&=\sqrt{ \xi_h^2+\psi_h^2}
\\
D_h&=\sqrt{x_h} dx_h
\\
\phi_h(n)&= -\sum_{j=1}^{N}
\left[\frac{\psi_j(n)\tanh\left({X_j/2\tau}\right)\upsilon^s(x_h,x_j)}{2X_j}\right]D_j
-\left[\frac{\psi_N(n)\tanh\left({X_N/2\tau}\right)\upsilon^s(x_h,x_N)}{2X_N}\right]\times\frac{2}{3}x_N^{3/2}
\label{eq:psi2phi}
\end{align}
The second term on the right hand side is the analytical integral in the interval $\{x_N;\infty\}$, taking into account that for $x_j\gg 1 $ both $\upsilon^s(x_i,x_j)\propto x_j^{-1}$ and $\psi_j(n)\propto x_j^{-1}$. For $x_j\gg 1 $ we can use $X_j\approx x_j$. Consequently the integrand is proportional to $x_j^{-5/2}$, from which follows the second term. 
The first and second derivatives of the grand potential are
\begin{align}
\frac{d w}{d \psi_i}&=
\frac{3}{4}
\left[
\frac{\xi_i^2\tanh\left({X_i}/{2\tau}\right)}{X_i^3}
+\frac{\psi_i^2}{2\tau X_i^2\cosh^2({X_i}/{2\tau})}
\right]
\left[\psi_i-\phi_i\right]D_i
\\
\frac{\partial^2 w}{\partial \psi_i^2}
&=
\frac{3}{4}
\frac{\psi_i}{X_i^3}
\left[
\frac{3\xi_i^2+\psi_i^2}{2\tau X_i^2\cosh^2\left({X_i}/{2\tau}\right)}
-\frac{\psi_i^2\tanh\left({X_i}/{2\tau}\right)}{4\tau^2\cosh^2\left({X_i}/{2\tau}\right)}
-\frac{3\xi_i^2\tanh\left({X_i}/{2\tau}\right)}{X_i^2}
\right]
\left[\psi_i-\phi_i\right]\delta_{i,m}D_i+
\nonumber\\
&\quad+
\frac{3}{4}\frac{1}{X_i^2}
\left[\frac{\xi_i^2 \tanh\left({X_i}/{2\tau}\right)}{X_i}
+\frac{\psi_i^2}{2\tau\cosh^2\left({X_i}/{2\tau}\right)}\right]
D_i+
\nonumber\\
&\quad+
\frac{3}{8}\frac{\upsilon^s(x_i,x_i)}{X_i^4} 
\left[\frac{\xi_i^2\tanh\left({X_i}/{2\tau}\right)}{X_i}
+
\frac{\psi_i^2}{2\tau\cosh^2\left({X_i}/{2\tau}\right)}\right]
\left[
\frac{x_{i}^2\tanh\left({X_{i}}/{2\tau}\right)}{X_{i}}
+
\frac{\psi_{i}^2}{2\tau\cosh^2\left({X_{i}}/{2\tau}\right)}
\right]
D_i^2
\end{align}
At the mininum  we have $d w / d \psi_i=0$ for all $\psi_i$. We search for the minimum using the steepest descent method, \textit{i.e.}, we iterate
\begin{equation}
\psi_i(n+1) = \psi_i(n) - \frac{\eta}{2}
\frac{\partial w(n)/\partial \psi_i(n)}
       {\partial^2 w(n)/\partial \psi_i(n)^2}
\end{equation}
where $\eta<1$ to ensure convergence of the iteration. 
The first step consisted of solving the equation for $r_s=50$, and this process was repeated down to $r_s=0.5$ using the output for $\psi_i$ as the starting values for each subsequent $r_s$ value. 
For the calculation of the temperature dependence, for each $r_s$ value the temperature was increased from 0 to $0.85\psi_F$ where $\psi_F$ is the order parameter at $\varepsilon_{\mathrm{F}}$, using the output for $\psi_i$ as the starting values for each subsequent temparature. 
Integrations were done using the trapezium method. The energy mesh used is specified in Table~1.

In each new iteration $\psi_F$ is taken to be the output value of previous iteration. Consequently, if the order parameter converges to a small value the set of energy values is adjusted correspondingly to a finer mesh around $\varepsilon_{\mathrm{F}}$.  Numerical verification indicates that $n=500$ is sufficiently large, giving a grand total of 6500 energy points at which the order parameter needs to be calculated self-consistently.

Although integrals over a log-divergence are convergent, jitter is unavoidable in the present iterative procedure due to finite sampling of the energy axis. To dampen the jitter we replace in Eq.~\ref{eq:upsilon_kernel}
\begin{equation}
\ln\left|\frac{4z^2-\left(\sqrt{x} -\sqrt{y}\right)^2 [4z_0^2- z^2\gamma^{-2}] }
{4z^2-\left(\sqrt{x} +\sqrt{y}\right)^2 [4z_0^2-z^2\gamma^{-2}]}\right|
\hspace{3mm}\mbox{with}\hspace{3mm}
\ln\left|\frac{4(z^2+i\delta^2)-\left(\sqrt{x} -\sqrt{y}\right)^2 [4z_0^2- z^2\gamma^{-2}] }
{4(z^2+i\delta^2)-\left(\sqrt{x} +\sqrt{y}\right)^2 [4z_0^2-z^2\gamma^{-2}]}\right|
\end{equation}
where $\delta = 2/n$. 
At equilibrium the grand potential becomes
\begin{equation}
w=
\frac{3}{4}\int_{0}^{\infty}  dx\sqrt{x}
\left[
-2\tau\ln\left(1+e^{-E_x/\tau}\right)
+\xi_x
-E_x  
+\frac{\psi_x^2}{2E_x}\tanh\left(\frac{E_x}{2\tau}\right)
\right]
\end{equation}
where $\psi_x$ is order parameter at the free energy minimum. For the purpose of displaying the energy dependence we choose the sign of $\psi_x$ to be positive at the Fermi surface.
The condensation energy is the energy saving at $T=0$ of the superconducting state compared to the trivial ($\vec{\psi}=0)$ solution of the gap-equation:
\begin{equation}
e_c=w_0^{(n)}-w_0^{(sc)}
\label{eq:econd}
\end{equation}

\subsection*{Solution without discretization}
\label{section:continuous}
The interaction $v^s_{x,y}$ entering the gap equation (\ref{eq:gap_equation}) presents logarithmic singularities. 
This may lead to inaccuracies when the integral is evaluated using a trapezoidal rule. 
We have investigated this issue by means of an alternative approach, where the gap function $\psi_x$ is represented as a piecewise-continuous function of $x$. In each piece, beside polynomials of arbitrary order, we use a variety of functions to capture the precise behavior of $\psi_x$, including logarithms and power laws. $\psi_y$ being a continuous function, the integral can be split at the singularities of $v^s_{x,y}$ and evaluated to the desired accuracy using Gaussian quadratures, without discretization error. At each iteration towards self-consistency, we fit our set of functions to $\psi_x$ in sub-domains that are recursively split until a satisfactory fit is found. This method is much slower, but produces results that are hardly distinguishable from those obtained by discretizing the integral (see Fig.~3). In Fig.~7 the difference of the two results is displayed. The differences are small, and most pronounced near the singularities. We conclude that our results are not significantly affected by discretization errors. 
\begin{figure}
\includegraphics[width=0.5\columnwidth]{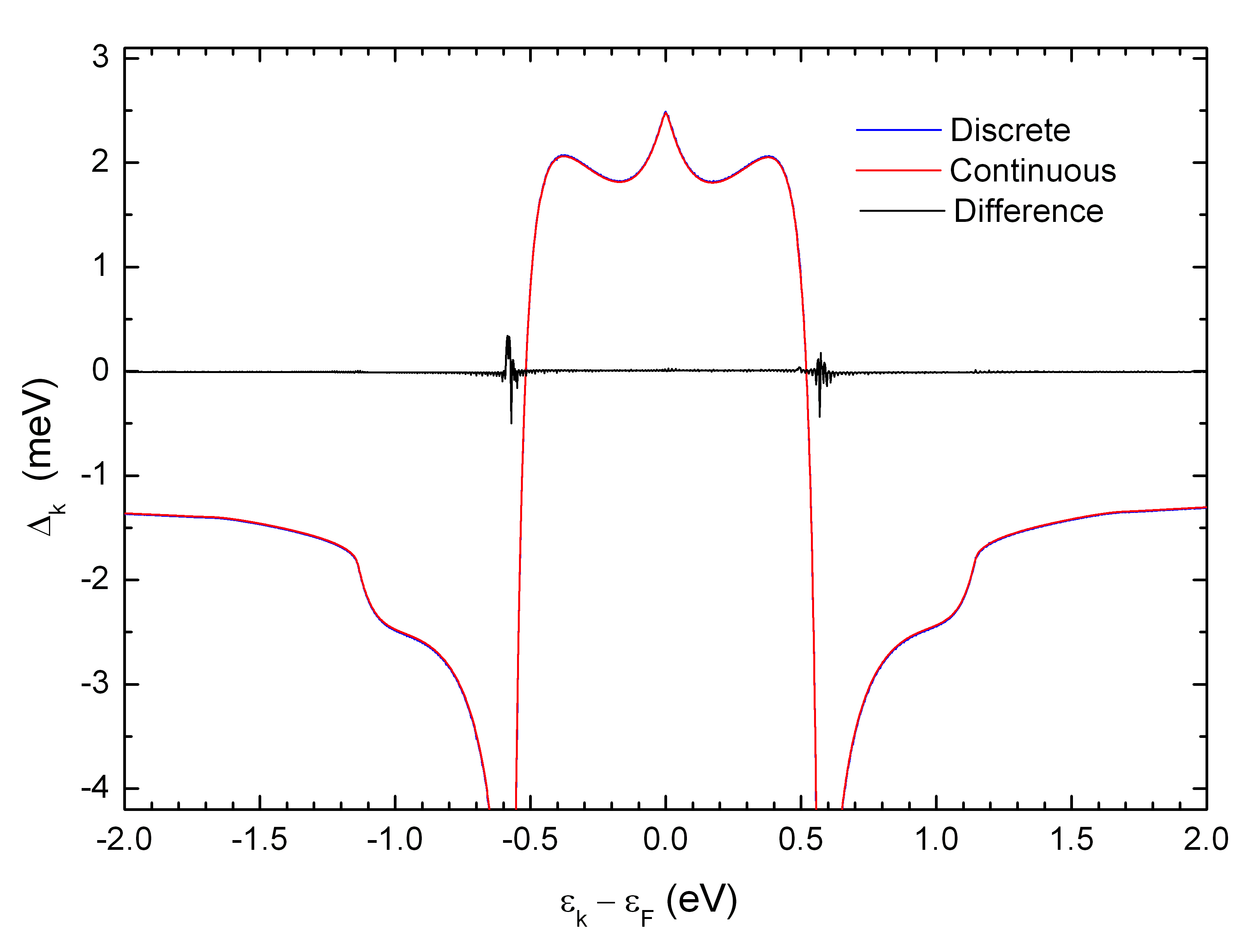}
\caption{Difference of the gap function at $T=0$ shown in Fig.~3. The results shown are calculated with two different numerical codes. $\Delta(\varepsilon)=\Delta_{\mathrm{cont}}(\varepsilon)-\Delta_{\mathrm{discr}}(\varepsilon)$, where the integrals in the numerical code for $\Delta_{\mathrm{discr}}(\varepsilon)$ are calculated with the trapezoidal rule (Methods - Numerical implementation), and those for $\Delta_{\mathrm{cont}}(\varepsilon)$ with a piecewise-continuous function of $x$ (Methods - Solution without discretization).
}
\end{figure}

\subsection*{Conventional approach}
\label{section:conventional}
Following DeGennes' approach~\cite{degennes1966} we obtained, from linear screening of the Coulomb interaction, Eq.~\ref{eq:1} and adopting the model dielectric function defined in Eq.~\ref{eq:2}, the effective electron-electron interaction Eq.~\ref{eq:Uscr2} in one fell swoop. A more common practice is to separate the Coulomb potential and the phonon-mediated interaction:
\begin{equation}
V^{\mathrm{eff}}(\vec{k},\vec{k}+\vec{q},\omega)=\frac{4\pi e^2}{k_0^2+q^2} 
+|M_{\vec{k},\vec{k}+\vec{q}}|^2\mathrm{Re}\,D(q,\omega),
\label{eq:split}
\end{equation}
where $M_{\vec{k},\vec{k}+\vec{q}}$ is the electron-phonon coupling function and $D(q,\omega)$ is the phonon propagator.
The conventional approach is to use for the electron-phonon coupling the Bardeen-Pines expression~\cite{bardeen1955} 
\begin{equation}
M_{\vec{k},\vec{k}+\vec{q}}=\frac{4\pi e^2 }{q\epsilon(q,0)} \sqrt{\frac{n}{2\omega_q m_N}},
\end{equation}
where
\begin{equation}
\epsilon(q,0)=1+\frac{k_0^2}{q^2}
\end{equation}
is the static dielectric constant describing Thomas-Fermi screening. The phonon propagator of the jellium model is
\begin{equation}
D(q,\omega) = \frac{1}{\omega+i0^+-\omega_q}-\frac{1}{\omega+i0^++\omega_q},
\label{eq:ph-jellium}
\end{equation}
where $\omega_q$ is the longitudinal acoustic phonon frequency. The momentum dependence follows from the zero's of the dielectric function, Eq.~\ref{eq:epsilon_q_omega}, with the result
\begin{equation}
\omega_q = \frac{q }{\sqrt{k_0^2+q^2}} \omega_0.
\end{equation}
The expression for $\omega_0$ is given in Eq.~\ref{eq:k0_omega0}.
Collecting all terms in Eq.~\ref{eq:split} provides the effective interaction
\begin{equation}
V^{\mathrm{eff}}(\vec{k},\vec{k}+\vec{q},\omega)  =
\frac{4 \pi e^2\omega^2}{k_{0}^2\omega^2 +(\omega^2-\omega_0^2)q^2} 
\end{equation}
which is the same expression as Eq.~\ref{eq:Uscr2}.

Usually the instantaneous part of the interaction, \textit{i.e.} the first term of Eq.~\ref{eq:split}, is approximated by it's average value at the Fermi level multiplied with the density of states $N_{\mathrm{F}}$
\begin{align}
\mu&=
N_{\mathrm{F}}
\sum_{\vec{k},\vec{q}} 
\frac{4\pi e^2}{k_0^2+q^2}
\frac{
\delta(\varepsilon_k-\varepsilon_{\mathrm{F}})
}{N_{\mathrm{F}}}
\frac{
\delta(\varepsilon_{|\vec{k}+\vec{q}|}-\varepsilon_{\mathrm{F}})
}{N_{\mathrm{F}}}
=\frac{\ln\left(1+\pi k_{\mathrm{F}} a_0 \right)}{2\pi k_{\mathrm{F}} a_0},
\label{eq:mu}
\end{align}
where in the last part the expression for the density of states at $\varepsilon_{\mathrm{F}}$
\begin{equation}
N_{\mathrm{F}}=\frac{m_e k_{\mathrm{F}}}{2\pi^2\hbar^2}
\end{equation}
was inserted.
Morel and Anderson~\cite{morel1962} demonstrated that, due to the retarded nature of the phonon-mediated interaction, the \emph{effective} Coulomb repulsion entering the expression for the gap is smaller than $\mu$ and is given by
\begin{equation}
\mu^*=\frac{\mu}{1+\mu\ln(\varepsilon_{\mathrm{F}}/\omega_0)}.
\end{equation}
The standard treatment of the phonon-mediated electron-electron interaction employs the electron-phonon coupling function~\cite{scalapino1965,zaccone2024}
\begin{equation}
\alpha^2F(\omega) = \sum_{\vec{k},\vec{q}} \frac{|M_{\vec{k},\vec{k}+\vec{q}}|^2}{N_{\mathrm{F}}}  \delta(\varepsilon_k-\varepsilon_{\mathrm{F}})\delta(\varepsilon_{|\vec{k}+\vec{q}|}-\varepsilon_{\mathrm{F}})A(q,\omega).
\end{equation}
The phonon spectral function corresponding to Eq.~\ref{eq:ph-jellium} is
\begin{equation}
A(q,\omega) = \delta(\omega-\omega_q)-\delta(\omega+\omega_q).
\end{equation}
We thus obtain for $\omega>0$
\begin{align}
\alpha^2F(\omega)  &=
\frac{8\pi^2 n e^4}{N_{\mathrm{F}} m_N}
\sum_{\vec{k},\vec{q}} 
\frac{q^2}{(k_0^2+q^2)^2}
\delta(\varepsilon_k-\varepsilon_{\mathrm{F}})\delta(\varepsilon_{|\vec{k}+\vec{q}|}-\varepsilon_{\mathrm{F}})
 \frac{\delta(\omega-\omega_q)}{\omega_q}.
\end{align}
We are now ready to calculate $\lambda$ using the standard expression 
\begin{align}
\lambda&=2\int_0^{\infty} \frac{\alpha^2F(\omega) }{\omega}d\omega
=
\frac{4\pi e^2}{N_{\mathrm{F}}}
\sum_{\vec{k},\vec{q}} 
\frac{1}{k_0^2+q^2}
\delta(\varepsilon_k-\varepsilon_{\mathrm{F}})\delta(\varepsilon_{|\vec{k}+\vec{q}|}-\varepsilon_{\mathrm{F}}),
\end{align}
where in the last part we recognize the result for $\mu$, Eq.~\ref{eq:mu}. We conclude that for the jellium model 
\begin{equation}
\lambda=\mu.
\end{equation}
We can now calculate $T_c$ and the isotope coefficient $-d\ln(T_c)/d\ln(m_N)$  using the McMillan equation~\cite{mcmillan1968} 
\begin{equation}
T_c=\frac{\omega_0}{1.45 k_B}\exp{\left\{-\frac{1.04(1+\lambda)}{\lambda-\mu^*(1+0.62\lambda)} \right\}}.
\label{eq:McMillan}
\end{equation}
The result, shown in Fig.~8, shows that $\lambda<0.5$, $\mu^*<0.25$, and $T_c < 25$ mK. Given that the coupling is weak, the small values of $T_c$ are not a surprise. Due to the fact that $\mu^*$ appearing in the exponent depends on the phonon-frequency $\omega_0$, there is a strong departure from the isotope effect corresponding to $\mu^*= 0$.
\begin{figure}
\includegraphics[width=0.5\columnwidth]{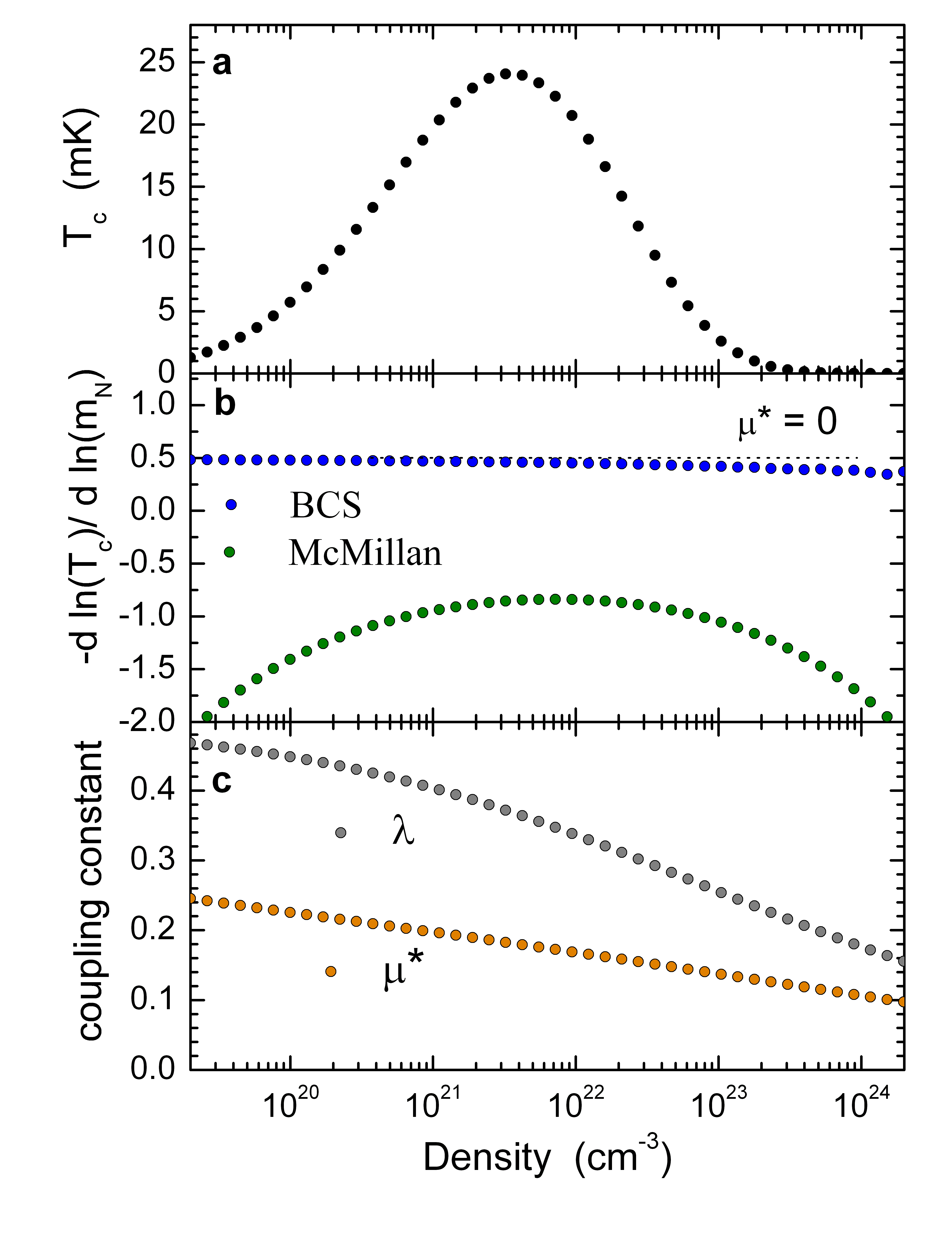}
\caption{Electron-phonon coupling constant $\lambda$, pseudopotential $\mu^*$, critical temperature and isotope coefficient calculated with the McMillan equation for $T_c$ with $m_N=m_p$ (hydrogen). In panel {\bf b} the points labeled “BCS” were calculated from the data shown in Fig.~5 using $\alpha=\ln (T_c^{\mathrm{H}}/T_c^{\mathrm{D}})/\ln(2)$ and the points labeled “McMillan” represent $- d\ln(T_c) / d ln(m_N)$ calculated with Eq.~\ref{eq:McMillan} and the parameters displayed in panel {\bf c}.
}
\end{figure}

\section*{Acknowledgments}
We gratefully acknowledge valuable comments of Peter Hirschfeld, Louk Rademaker, Jorge Hirsch, George Sawatzky, Thierry Courvoisier and the reviewers of this manuscript. 

\section*{Author contributions}
DvdM conceived the project. DvdM and CB developed the theoretical expressions, wrote the numerical code and performed the calculations. DvdM wrote the manuscript with input from CB.

\section*{Declaration of interests}
The authors declare no competing interests.

\bigskip

%


%
\end{document}